# Scintillation index analysis for multi-wavelength Gaussian beams in turbulent underwater channels

**Shideh Tayebnaimi*** **and Kamran Kiasaleh**
The University of Texas at Dallas, Erik Jonsson School of Engineering and Computer Science, Richardson, Texas, United States

**ABSTRACT.** We investigate the scintillation index of multi-wavelength Gaussian optical beams propagating through a turbulent optical channel. We consider a turbulent environment ranging from weak to strong, with a specific focus on the weak turbulent regime. Furthermore, the impacts of absorption and scattering effects are taken into account. It is shown here that the use of a multi-wavelength beam results in a reduction in the scintillation index, thereby enhancing the performance of the underwater optical link.

 

## 1 Introduction

In recent years, the demand for underwater communication applications has steadily increased due to applications such as collecting scientific data, monitoring the environment, exploring oil fields, and shipping to submerged platforms (submarines), maritime archaeology, and port security, which demand the ability to communicate in an underwater environment. These applications have fueled the need for high-speed wireless communication and high-quality imaging in underwater environments.[1] Furthermore, novel applications such as the Internet of Underwater Things, network-centric communication and imaging, and communication among autonomous underwater vehicles, along with underwater-to-satellite communication and underwater-to-ground, have attracted substantial interest from researchers in recent decades. However, underwater wireless optical communications (UWOC) face limitations in link length, typically spanning tens of meters. This constraint results from the interplay of turbulence phenomena, scattering, and absorption, driven by the chaotic and challenging conditions of underwater media. Among these, scattering and absorption play dominant roles in attenuating optical waves in the underwater environment.[2–4] Although optical signaling holds promise for achieving high-quality imaging underwater and high-speed wireless connectivity, it faces significant challenges due to optical turbulence in water.

Turbulence occurs due to rapid, though spatially gentle, fluctuations in the refractive index of the water. To address this, precise analytical models of the spatial power spectrum across various conditions are required. These conditions are the result of temperature and salinity fluctuations, which are the two primary factors impacting optical turbulence. In turbulent water, the refractive index power spectrum is significantly influenced by the average fluctuations in temperature and salinity and, consequently, optical signal transmission.[5,6]

*Address all correspondence to Shideh Tayebnaimi, shideh.tayebnaimi@utdallas.edu

Researchers investigate UWOC channel models by accounting for various noise sources utilizing radiative transfer theory.[7] The scintillation index, which reflects the resulting intensity fluctuations, emphasizes the considerable influence of turbulence on optical signal transmission.[8,9]

The majority of investigations into laser beam propagation center around single-wavelength lasers encountering optical turbulence. To reduce the impact of turbulence, employing a diversity measure is suggested. One approach involves exploiting the wavelength-specific characteristics of the turbulent medium.[10–15] Specifically, in Ref. 10, it has been shown that the scintillation index of an optical beam with multiple wavelengths is superior to that of a single-wavelength beam.

In contrast to Free Space Optics (FSO), UWOC experiences significantly greater propagation losses due to several key factors, including the variability of the refractive index with changes in water conditions and the limited usable power spectrum affected by selective absorption of light. In general, there are two unique characteristics of UWOC that set this analysis apart from its FSO counterpart.

- Refractive index and power spectrum: The refractive index in FSO is generally stable and minimally affected by environmental changes, whereas in UWOC, it varies significantly with changes in water temperature, salinity, and pressure, introducing refraction effects that can distort optical signals.
- Signal attenuation and absorption: In FSO, signal attenuation primarily depends on distance and atmospheric conditions, such as turbulence, fog, or rain, which scatter light and reduce visibility. UWOC suffers from much more pronounced signal degradation due to absorption and scattering from water molecules and suspended particles, as well as turbulence caused by water currents or thermal gradients, leading to rapid attenuation that exceeds that of FSO systems.

## 2 Channel and System Models

The analysis assumes that backscattering and depolarization effects are negligible, the refractive index is delta-correlated along the direction of propagation, and the parabolic (paraxial) approximation is applicable. By assuming a sinusoidal time variation (i.e., a monochromatic wave) in the electric field, it is well established that Maxwell's equations for the vector amplitude $E(R)$ of a propagating electromagnetic wave directly result in:[16]

$$\nabla^2 E(R) + \kappa^2 n^2(R) E(R) = 0, \tag{1}$$

where $R$ represents a vector in the transversal plane, $E(R)$ denotes the electric field magnitude of the light, $k = \frac{2\pi}{\lambda}$ is the wave number of the electromagnetic wave, $\lambda$ is the wavelength, $n(R)$ is the refractive index with the suppressed time variations, and $\nabla^2$ is the Laplacian operator. This concept can be extended to the multi-wavelength light case (for more details, see Appendix A in Ref. 11). In that event, we have

$$\nabla^2 E_l(R) + \kappa_l^2 n^2(R) E_l(R) = 0; \qquad l = 1, \ldots, M, \tag{2}$$

where the wave number is defined as $k_l = \frac{2\pi}{\lambda_l}$ for the $l$'th wavelength and $E_l(R)$ represents the magnitude of the electric field for the $l$'th wavelength of the light.

The electric field produced by the multi-wavelength optical radiation is described as

$$E(R,t) = \sum_{l=1}^{M} E_l(R) \exp\left(j\left(\frac{2\pi ct}{\lambda_l} + \Phi_l(t)\right)\right), \tag{3}$$

with $E(R,t)$ is the electric field at location $R$ and time $t$ and $\Phi_l(t)$ refers to the phase noise associated with the $l$'th wavelength, causing the $l$'th wavelength component of the beam to exhibit a Lorentzian spectral shape. By applying the Rytov approximation, a perturbation method for modeling the received optical beam yields

$$E^{(r)}(R,t) = \sum_{l=1}^{M} E_l^{(r)}(R)\exp(\Psi_l(R))\exp\left(j\left(\frac{2\pi ct}{\lambda_l} + \Phi_l(t)\right)\right), \tag{4}$$

in which $E_l^{(r)}(R)$ represents the received electric field for the $l$'th mode without the atmospheric turbulence and $\Psi_l(R)$ indicates the complex amplitude and phase perturbation caused by turbulence.

## 2.1 Power Spectrum

The power spectrum of a Gaussian beam based on the modified Nikishov spectrum can be represented as follows:[17]

$$\Phi_n(\kappa) = \frac{1}{4\pi} C_0 \left(\frac{\alpha^2 \chi_T}{\omega^2}\right) \epsilon^{-\frac{1}{3}} \kappa^{-\frac{11}{3}} [1 + C_1(\kappa\eta)^{\frac{2}{3}}] \times \Big[\omega^2 \exp(C_0 C_1^{-2} P_T^{-1}\delta) + d_r \exp(C_0 C_1^{-2} P_S^{-1}\delta) - \omega(d_r + 1)\exp\left(\frac{-C_0 C_1^{-2} P_{TS}^{-1}}{2}\delta\right)\Big], \tag{5}$$

with $\kappa$ representing the magnitude of the spatial frequency and $\varepsilon$ (measured in $m^2/s^3$) denotes the energy dissipation rate. The constants $C_0$ and $C_1$ have values of 0.72 and 2.35, respectively. In addition, $\chi_T$ is the thermal expansion coefficient of the Kolmogorov micro-scale length and $\eta$ (in $m^{-1}$) is given by $\eta = \nu^{\frac{4}{3}}\varepsilon^{-\frac{1}{4}}$, where $\nu$ is the kinematic viscosity. $\omega$ represents the relative intensity of temperature and salinity fluctuations and can be determined as

$$\omega = \frac{\alpha(\mathrm{dT}/\mathrm{dz})}{\beta(\mathrm{dS}/\mathrm{dz})}, \tag{6}$$

where $\alpha$ and $\beta$ are the coefficients for thermal expansion and saline contraction, respectively. Moreover, $(dT/dz)$ represents the temperature difference, and $(dS/dz)$ denotes the salinity difference between the top and bottom boundaries.

In Eq. (5), the Prandtl number for temperature is denoted by $P_T$, whereas $P_S$ represents the Prandtl number for salinity. $P_{\mathrm{TS}}$ is defined as half the harmonic mean of $P_T$ and $P_S$, $d_r$ stands for the eddy diffusivity ratio, and $\delta = 1.5c_1^2(\kappa\eta)^{\frac{4}{3}} + C_1^3(\kappa\eta)^2$.[18,19]

## 2.2 Scintillation Index

Scintillation, which refers to fluctuations in received irradiance caused by turbulence during the propagation of optical waves, can be analyzed by considering the scintillation index varying with the transverse vector length $r = |R|$ on the transverse plane and the propagation distance $L$. The dependency on $r$ is due to the symmetrical nature of the Gaussian beam in the transversal plane. The scintillation index can therefore be expressed as[16,20]

$$\sigma_I^2(r,L) = \frac{\left\langle I^2(r,L)\right\rangle}{\left\langle I(r,L)\right\rangle^2} - 1, \tag{7}$$

where $I(r,L)$ is the intensity of the received signal and $\langle\rangle$ denotes an ensemble average of the enclosed. To assess the scintillation index, we first compute the irradiance of the received optical field. The intensity of the received signal can be represented by the following model:

$$I(r,L) = \langle|E^{(r)}(r,L)|^2\rangle_t = \sum_{l=1}^{M} |E_l^{(r)}(r,L)|^2 \exp[\psi_l(r,L) + \psi_l^*(r,L)], \tag{8}$$

where $\langle\rangle_t$ signifies the time average of the enclosed quantity. Given that $|E_l^{(r)}(r,L)|^2 = A_l(r,L)$ denotes the signal intensity for the $l$'th beam component in a non-turbulent scenario, the intensity of the multi-wavelength received field is given by

$$I(r,L) = \sum_{l=1}^{M} A_l(r,L)\exp[\psi_l(r,L) + \psi_l^*(r,L)]. \tag{9}$$

Letting $\zeta_l(r,L) = \mathrm{Re}[\psi_l(r,L)]$ denote the real part of $\psi_l(r,L)$, the total intensity $I(r,L)$ can be expressed as

$$I(r,L) = \sum_{l=1}^{M} A_l(r,L)\exp[2\zeta_l(r,L)]. \tag{10}$$

To further quantify the average intensity, $\overline{I(r,L)}$, we utilize the same structure and express it as

$$\overline{I(r,L)} = \sum_{l=1}^{M} A_l(r,L)\overline{\exp[2\zeta_l(r,L)]}, \tag{11}$$

where $\zeta_l(r,L)$ denotes the random component associated with each mode. By inserting the expressions for $I(r,L)$ and $\overline{I(r,L)}$ from Eqs. (10) and (11) into the definition of the scintillation index from (7), we obtain

$$\sigma_{\mathrm{sc}}^2 = \frac{\sum_{l_1=1}^{M}\sum_{l_2=1}^{M} A_{l_1}(r,L)A_{l_2}(r,L)\overline{(\exp[2\zeta_{l_1}(r,L) + 2\zeta_{l_2}(r,L)])}}{[\sum_{l=1}^{M} A_l(r,L)\overline{\exp[2\zeta_l(r,L)]}]^2} - 1. \tag{12}$$

It follows that $\Gamma_l = \overline{\exp[2\zeta_l(r,L)]}$ and $\Gamma_{l_1 l_2} = \overline{\exp[2\zeta_{l_1}(r,L) + 2\zeta_{l_2}(r,L)]}$ are required to continue.

Thus, $\gamma_l(r,L)$ is defined as the mean of the real part of $\psi_l(r,L)$, as shown by $\gamma_l(r,L) = \overline{\zeta_l(r,L)} = \overline{\mathrm{Re}[\psi_l(r,L)]}$. Consequently, we can express the above equations as shown below:

$$\Gamma_l(r,L) = \overline{\exp[2\zeta_l(r,L)]} = \exp[2(\gamma_l(r,L) + \sigma_l^2(r,L))], \tag{13}$$

$$\overline{I(r,L)} = \sum_{l=1}^{M} A_l(r,L)\Gamma_l, \tag{14}$$

$$\Gamma_{l_1 l_2}(r,L) = \overline{\exp[2\zeta_{l_1}(r,L) + 2\zeta_{l_2}(r,L)]} = \Gamma_{l_1}\Gamma_{l_2}\exp[4R_{l_1 l_2}(r,L;r,L)], \tag{15}$$

and

$$\sigma_{\mathrm{sc}}^2 = \frac{\sum_{l_1=1}^{M}\sum_{l_2=1}^{M} A_{l_1}(r,L)A_{l_2}(r,L)\Gamma_{l_1 l_2}(r,L)}{[\sum_{l=1}^{M} A_l(r,L)\Gamma_l(r,L)]^2} - 1. \tag{16}$$

To calculate $\sigma_{\mathrm{sc}}^2$, it is evident that one needs not only the variance of $\zeta_l(r,L)$, given by

$$\begin{aligned} R_{ll}(\mathbf{r},\mathbf{L};\mathbf{r},\mathbf{L}) = \sigma_l^2(\mathbf{r},\mathbf{L}) = \overline{\zeta_l^2(\mathbf{r},\mathbf{L})} &= \frac{1}{2}\mathrm{Re}[\overline{\Psi_l(\mathbf{r},\mathbf{L})\Psi_l^*(\mathbf{r},\mathbf{L})}] \\ &\quad + \frac{1}{2}\mathrm{Re}[\overline{\Psi_l(\mathbf{r},\mathbf{L})\Psi_l(\mathbf{r},\mathbf{L})}], \end{aligned} \tag{17}$$

but also the correlation between $\zeta_{l_1}(r,L)$ and $\zeta_{l_2}(r,L)$, expressed as

$$\begin{aligned} R_{l_1 l_2}(\mathbf{r},\mathbf{L};\mathbf{r},\mathbf{L}) = \overline{\zeta_{l_1}(\mathbf{r},\mathbf{L})\zeta_{l_2}(\mathbf{r},\mathbf{L})} &= \frac{1}{2}\mathrm{Re}[\overline{\Psi_{l_1}(\mathbf{r},\mathbf{L})\Psi_{l_2}^*(\mathbf{r},\mathbf{L})}] \\ &\quad + \frac{1}{2}\mathrm{Re}[\overline{\Psi_{l_1}(\mathbf{r},\mathbf{L})\Psi_{l_2}(\mathbf{r},\mathbf{L})}];\ l_1 \neq l_2. \end{aligned} \tag{18}$$

By observing that $R_{l_1 l_2}(r,L;r,L) = R_{l_2 l_1}(r,L;r,L)$ and $\Gamma_{l_1 l_2}(r,L) = \Gamma_{l_2 l_1}(r,L)$, after solving and combining the equations, the scintillation index of a multi-wavelength beam can be computed as follows:[11]

$$\sigma_{sc}^2(r,L)=\frac{\sum_{l=1}^{M}A_l^2(r,L)\Gamma_l^2(r,L)\exp(4\sigma_l^2(r,L))+2\sum_{l_1=1}^{M}\sum_{l_2=1}^{l_1-1}A_{l_1}(r,L)A_{l_2}(r,L)\Gamma_{l_1l_2}(r,L)}{[\sum_{l=1}^{M}A_l(r,L)\Gamma_l(r,L)]^2}-1, \tag{19}$$

where $\Gamma_l(r,L)$ and $\Gamma_{l_1l_2}(r,L)$ are given by Eqs. (13) and (15), respectively.

Further, for a Gaussian beam, $\sigma_l^2(r,L)$, $\gamma_l(r,L)$, and $R_{l1l2}(r,L;r,L)$ can be represented as follows[11,16] [pp. 145-151]:

$$\sigma_l^2(r,L)=2\pi^2k_l^2L\int_0^1\int_0^\infty \kappa\Phi_n(\kappa)\exp\left(\frac{-\kappa^2\eta^2L\Lambda_l}{k_l}\right) \times I_0\left\{(2\kappa\Lambda_l|r|)-\left(\cos\left(\frac{\kappa^2L\eta(1-\eta)\hat{\Theta}_l}{k_l}\right)\right)\right\}\mathrm{d}\kappa\mathrm{d}\eta, \tag{20}$$

$$\gamma_l(r,L)=-2\pi^2k_l^2L\int_0^1\int_0^\infty \kappa\Phi_n(\kappa) \times\left\{1-\exp\left(\frac{-\kappa^2\eta^2L\Lambda_l}{k_l}\right)\cos\left(\frac{\kappa^2L\eta(1-\eta)\hat{\Theta}_l}{k_l}\right)\right\}\mathrm{d}\kappa\mathrm{d}\eta, \tag{21}$$

and

$$\begin{aligned}R_{l_1l_2}(r,L;r,L)=2\pi^2k_{l_1}k_{l_2}L\int_0^1\int_0^\infty \kappa\Phi_n(\kappa)\\ \times\mathrm{Re}\Bigg\{J_0(\kappa\eta([\Theta_{l_1}-\Theta_{l_2}]-j(\Lambda_{l_1}+\Lambda_{l_2}))|r|)\\ \times\exp\left[-j\kappa^2L\left(\frac{(1-\eta)+\eta(\Theta_{l_1}-j\Lambda_{l_1})}{2k_{l_1}}-\frac{(1-\eta)+\eta(\Theta_{l_2}+j\Lambda_{l_2})}{2k_{l_2}}\right)\eta\right]\\ -J_0(\kappa\eta([\Theta_{l_1}-\Theta_{l_2}]-j(\Lambda_{l_1}-\Lambda_{l_2}))|r|)\\ \times\exp\left[-j\kappa^2L\left(\frac{(1-\eta)+\eta(\Theta_{l_1}-j\Lambda_{l_1})}{2k_{l_1}}+\frac{(1-\eta)+\eta(\Theta_{l_2}-j\Lambda_{l_2})}{2k_{l_2}}\right)\eta\right]\Bigg\}\mathrm{d}\kappa\mathrm{d}\eta,\end{aligned} \tag{22}$$

respectively. In the above, $I_0(x)=J_0(jx)$ is the modified Bessel function of the first kind and order zero, and $J_0()$ is the Bessel function of zero order. Also, the Fresnel ratio of the Gaussian beam and the curvature parameter at receiver planes are $\Lambda_l=\Lambda_{0,l}/(\Theta_0^2+\Lambda_{0,l}^2)$ and $\Theta_l=\Theta_0/(\Theta_0^2+\Lambda_{0,l}^2)$, respectively. Moreover, the Fresnel ratio and the curvature parameter at the transmitter planes are $\Lambda_0=2\ L/k_lW_0^2$ and $\Theta_0=1-(L/F_0)$, respectively, where $W_0$ is the beam radius, $F_0$ is the phase front radius of curvature, and $\hat{\Theta}_l=1-\Theta_l$ is the complementary parameter.

## 3 Numerical Results

The numerical analysis is carried out, and without loss of generality, several parameters are fixed. That is, the temperature is set at 20°C and salinity at 35 ppt. Pressure = 0 dBar and $\omega=-0.3508$.[21] Furthermore, a collimated beam ($F_0=\infty$) and the scintillation index are evaluated at the beam's focal point, i.e., $r=0$. Given that absorption holds the greatest influence in underwater environments, the segment of the visible light spectrum ranging from ~450 to 485 nm, known as the blue region, experiences minimal attenuation in comparison to different areas of the light spectrum.[22] To underscore the overall advantage of wavelength diversity, various wavelengths spanning from 480 to 600 nm are employed across different regions, including the blue-green, green, and yellow regions.

Figure 1 illustrates how the scintillation index $\sigma_{sc}^2(r,L)$ varies with the propagation distance $L$, for the case where $A_l(r,L)=1$. The scintillation index is calculated for each individual wavelength: $M=1$ ($\lambda=480$ nm, $\lambda=532$ nm, and $\lambda=600$ nm), pairs of wavelengths $M=2$ (($\lambda=480$ and $\lambda=532$ nm), ($\lambda=480$ and $\lambda=600$ nm), and ($\lambda=532$ and $\lambda=600$ nm)), and for $M = 3$ ($\lambda = 480$, $\lambda = 532$, and $\lambda = 600$ nm) to demonstrate the advantages of wavelength diversity. The chosen wavelength range (480 to 600 nm) is limited due to absorption characteristics in underwater environments.

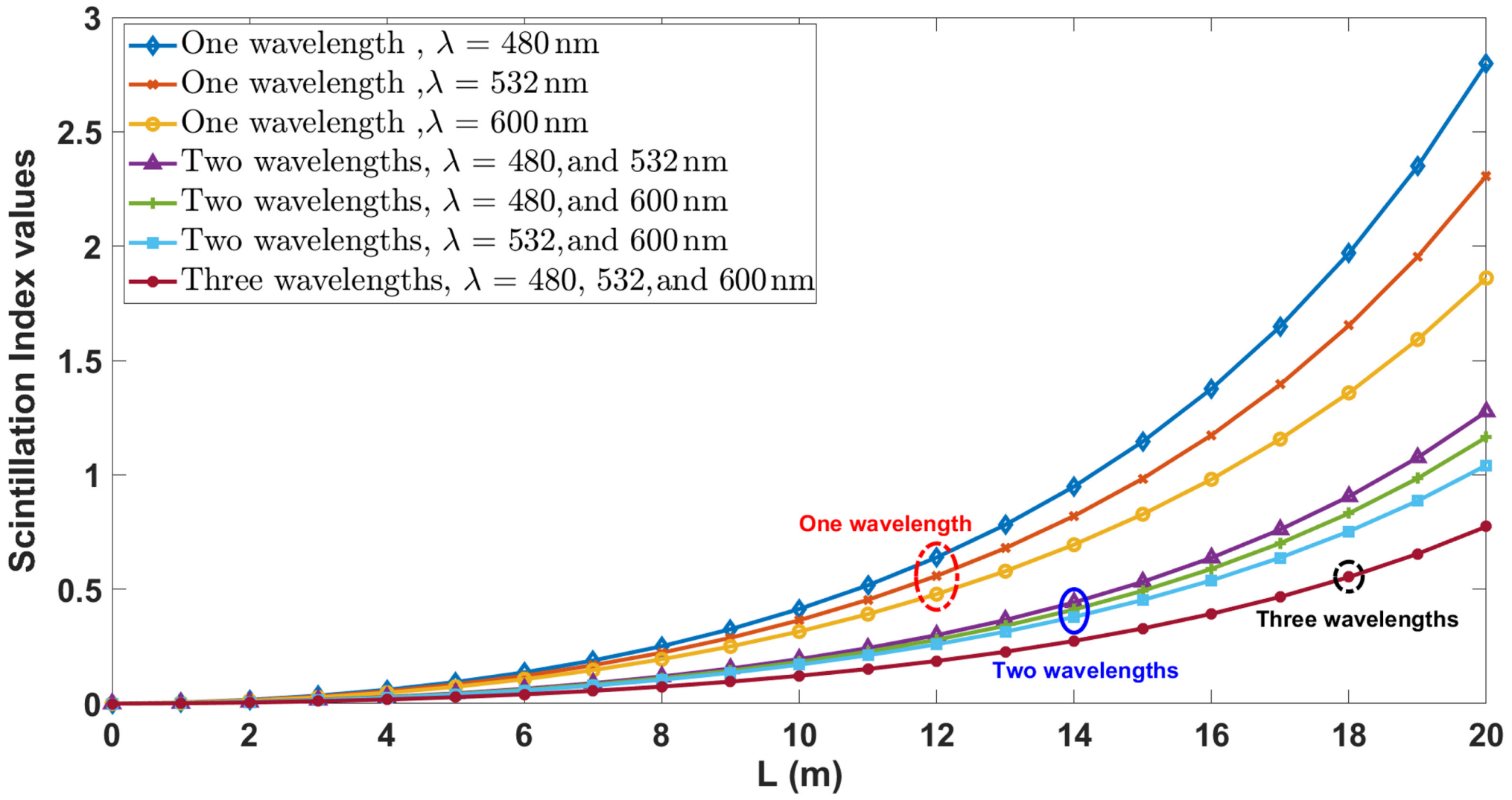

**Fig. 1** Scintillation index versus propagation distance.

The data clearly indicate that wavelength diversity significantly influences the scintillation index at the center of the beam for the wavelengths investigated. For instance, at a typical distance of $L = 10$ m, using two wavelengths results in a reduction of at least 42%, whereas employing three wavelengths achieves a reduction of up to 60% compared with using a single wavelength. In addition, switching from two to three wavelengths results in a further reduction of the scintillation index exceeding 28%.

With increasing the propagation distance (e.g., $L = 20$ m), the impact of wavelength diversity becomes even more pronounced. These findings underscore the significant benefits of employing multiple wavelengths to mitigate scintillation effects, especially over longer distances.

In Fig. 2, the scintillation index is plotted against $W_0$, which represents the beam radius at the transmitter. As anticipated, there is a slight increase in the scintillation index with higher values of $W_0$, particularly at higher propagation distances. Interestingly, the scintillation index shows a steady increase with longer propagation distances and remains relatively constant over these distances. Moreover, this illustrates how utilizing multiple wavelengths can effectively reduce scintillation at long distances. This method is particularly advantageous for maintaining signal quality across varying distances and beam conditions.

Figure 3 demonstrates an increase in the scintillation index as the temperature dissipation rate changes. This occurs because higher dissipation rates lead to a greater energy transfer from larger turbulent eddies to smaller ones, resulting in more pronounced velocity fluctuations. In turbulent flows, the chaotic variation of fluid velocity, caused by the presence of eddies and vortices of varying sizes, directly impacts intensity fluctuations. The comparison between Figs. 3(a) and 3(b) highlights that the scintillation index increases with rising temperature and higher energy dissipation rates, emphasizing the influence of temperature fluctuations on scintillation behavior in underwater environments. This is because the viscosity of water decreases with increasing temperature, allowing it to flow more easily and resist shear flow less. Lower viscosity at higher temperatures enhances the formation and size of turbulent eddies. In regions with lower viscosity, energy from turbulent eddies cascades more effectively to smaller scales, where it dissipates as heat, leading to a higher energy dissipation rate in warmer water.

Figure 4 demonstrates how the scintillation index is affected by the energy dissipation rate at $L = 10$ m. The energy dissipation rate ($\epsilon$) is a critical parameter reflecting the intensity of turbulent mixing in the underwater environment. This turbulence directly influences optical wave propagation through fluctuations in the water's refractive index. As $\epsilon$ increases, a notable decrease in the scintillation index is observed across all wavelength configurations, including both single and multiple wavelengths. This inverse relationship between $\epsilon$ and the scintillation index can be attributed to the nature of turbulent mixing. Higher $\epsilon$ values indicate more intense turbulence, which tends to homogenize the water medium by reducing the scale of refractive index fluctuations. Consequently, the optical wavefront experiences less severe phase and amplitude distortions, leading to a reduction in scintillation effects.

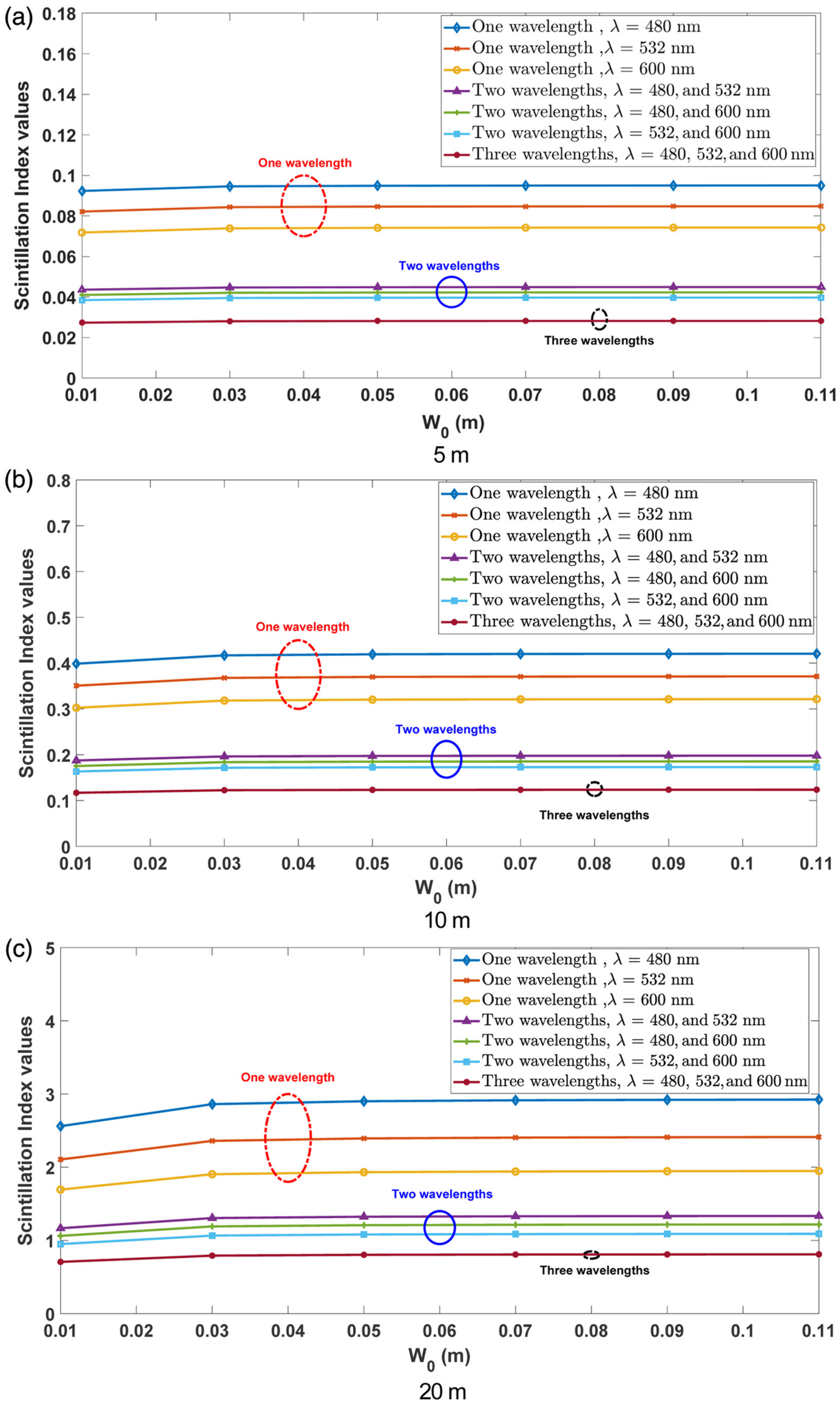

**Fig. 2** Scintillation index versus transmitter beam radius at various distances.

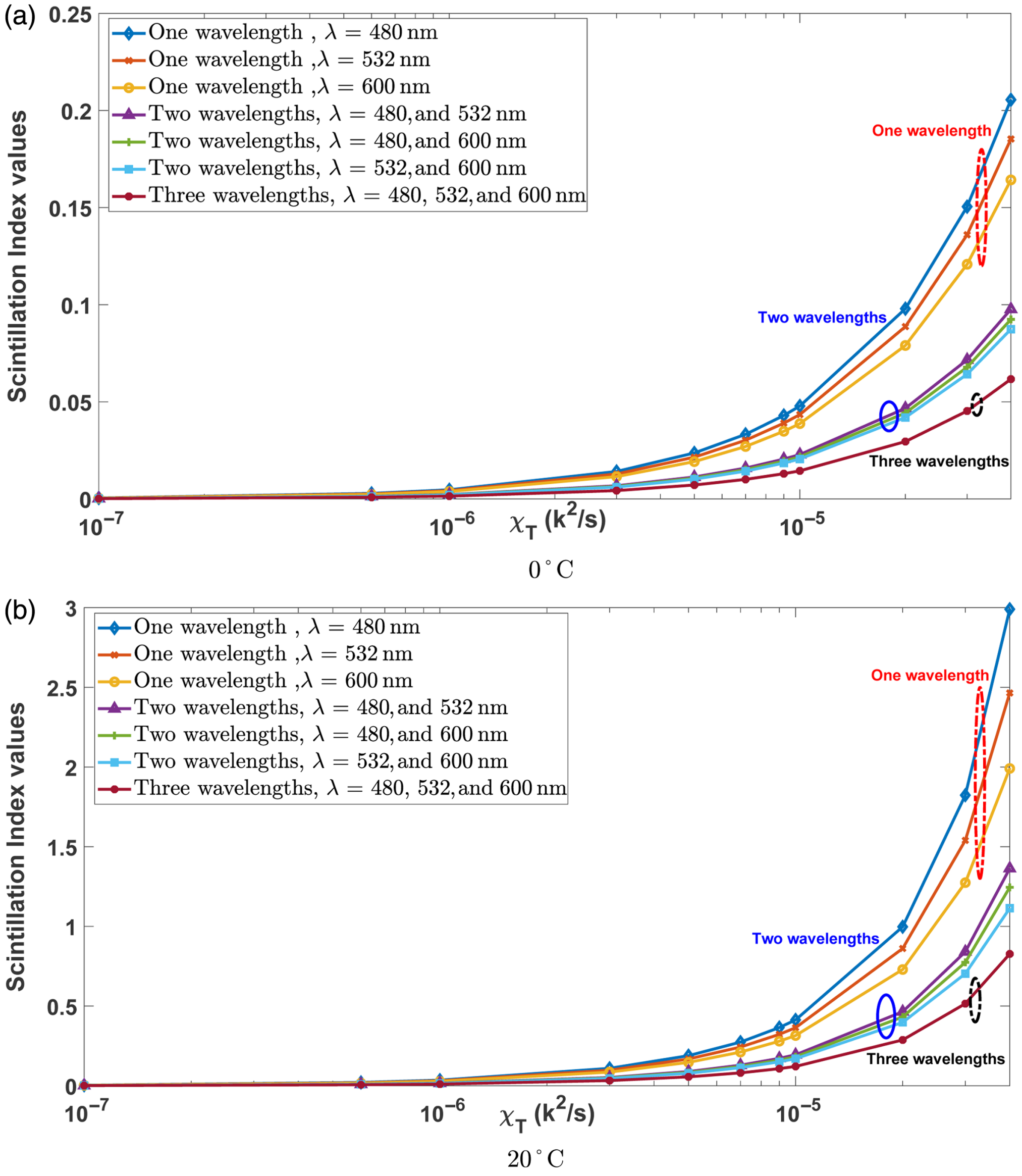

**Fig. 3** Scintillation index versus temperature dissipation rate at various temperatures.

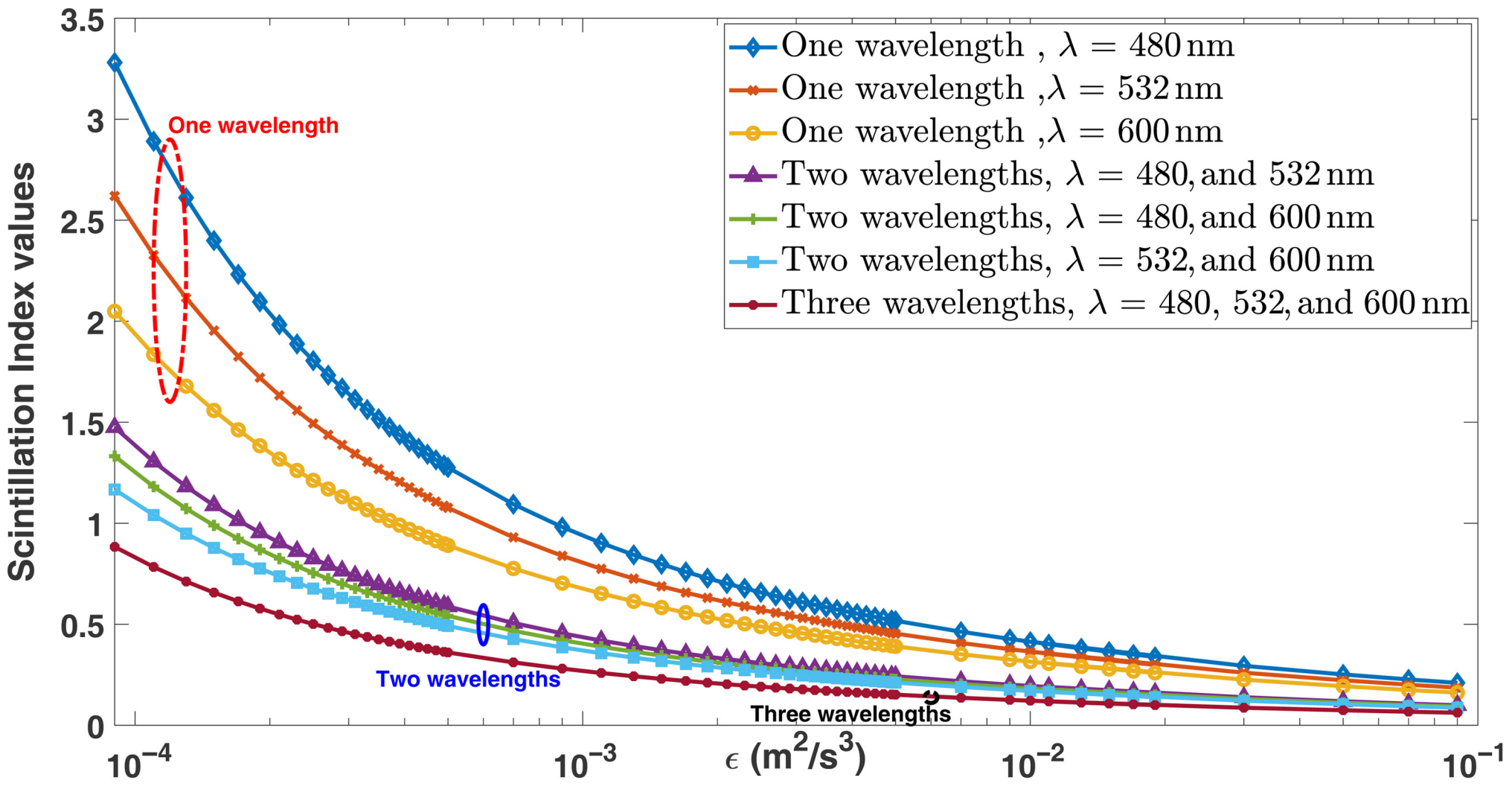

**Fig. 4** Scintillation index versus energy dissipation rate.

The analysis reveals that single-wavelength systems are more susceptible to scintillation, particularly at shorter wavelengths. For instance, the shortest wavelength ($\lambda = 480$ nm) consistently exhibits the highest scintillation index values across all energy dissipation rates, due to its increased sensitivity to smaller-scale refractive index variations. By contrast, longer wavelengths ($\lambda = 600$ nm) show comparatively lower scintillation index values, as they are less affected by these small-scale fluctuations. However, the deployment of multi-wavelength systems significantly mitigates scintillation effects. The data demonstrate that using two or three wavelengths leads to a pronounced reduction in the scintillation index, with the three-wavelength configuration ($\lambda = 480, 532$, and 600 nm) consistently achieving the lowest scintillation index values. This reduction can be attributed to the diversity effect, where different wavelengths interact with the turbulent medium in a complementary manner, effectively averaging out the refractive index fluctuations over the spectrum of wavelengths.

## 4 Conclusion

In this study, a wavelength diversity method was presented to reduce the scintillation impact. A multi-wavelength beam was studied for practical implementation, where beams of distinct wavelengths are merged at the transmitter and then sent along a singular optical route. The study investigated several parameters, including temperature, salinity concentration, link length, wavelength, and energy dissipation rates. These factors play a crucial role in understanding and optimizing underwater optical communication systems. Numerical analysis of the presented results highlights the impact of increasing the number of wavelengths on the deduction of the scintillation index. Another significant finding demonstrates the impact of temperature on the dissipation rate and, consequently, the scintillation index due to thermal eddies.

---

### Disclosures

The authors declare that there are no financial interests, commercial affiliations, or other potential conflicts of interest that could have influenced the objectivity of this research or the writing of this paper.

### Code and Data Availability

The archived version of the code described in this manuscript can be freely accessed through Code Ocean.

### References

1. O. Korotkova, “Light propagation in a turbulent ocean,” *Progr. Opt.* **64**, 1–43 (2018).
2. C. D. Mobley, “Comparison of numerical models for computing underwater light fields,” *Appl. Opt.* **32**(36), 7484–7504 (1993).
3. C. D. Mobley, *Light and Water: Radiative Transfer in Natural Waters*, Academic Press, San Diego, CA, USA (1994).
4. A. Morel and L. Prieur, “Analysis of variations in ocean color,” *Limnol. Oceanogr.* **22**, 709–722 (1977).
5. X. Quan and E. S. Fry, “Empirical equation for the index of refraction of seawater,” *Appl. Opt.* **34**, 3477–3480 (1995).
6. M. Kelly, S. Avramov-Zamurovic, and C. Nelson, “Exploration of multiple wavelength laser beams propagating underwater,” *Proc. SPIE* **10631**, 1063118 (2018).
7. S. Jaruwatanadilok, “Underwater wireless optical communication channel modeling and performance evaluation using vector radiative transfer theory,” *IEEE J. Sel. Areas Commun.* **26**, 1620–1627 (2008).
8. O. Korotkova, N. Farwell, and E. Shchepakina, “Light scintillation in oceanic turbulence,” *Waves Random Complex Media* **22**, 260–266 (2012).
9. Y. Ata and Y. Baykal, “Scintillations of optical plane and spherical waves in underwater turbulence,” *J. Opt. Soc. Amer. A* **31**, 1552–1556 (2014).

10. K. Kiasaleh, "Scintillation index of a multiwavelength beam in turbulent atmosphere," *J. Opt. Soc. Amer. A* **21**, 1452–1454 (2004).
11. K. Kiasaleh, "On the scintillation index of a multiwavelength gaussian beam in a turbulent free-space optical communications channel," *J. Opt. Soc. Amer. A* **23**, 557–566 (2006).
12. M. Moghaddasi et al., "Development of SAC-OCDMA in FSO with multiwavelength laser source," *Opt. Commun.* **356**, 282–289 (2015).
13. R. Purvinskis et al., "Multiple-wavelength free-space laser communications," *Proc. SPIE* **4975**, 12–19 (2003).
14. P. Corrigan et al., "Quantum cascade lasers and the Kruse model in free space optical communication," *Opt. Express* **17**(6), 4355–4359 (2009).
15. K. A. Balaji and K. Prabu, "Performance evaluation of FSO system using wavelength and time diversity over Malaga turbulence channel with pointing errors," *Opt. Commun.* **410**, 643–651 (2018).
16. L. C. Andrews and R. L. Phillips, *Laser Beam Propagation through Random Media*, 2nd ed., SPIE Press, Bellingham, Washington (2005).
17. M. Elamassie et al., "Effect of eddy diffusivity ratio on underwater optical scintillation index," *Phys. Oceanogr.* **34**, 1969–1973 (2017).
18. P. R. Jackson and C. R. Rehmann, "Laboratory measurements of differential diffusion in a diffusively stable, turbulent flow," *J. Opt. Soc. Amer. A* **33**, 1592–1603 (2003).
19. J. Yao, M. Elamassie, and O. Korotkova, "Spatial power spectrum of natural water turbulence with any average temperature, salinity concentration, and light wavelength," *J. Opt. Soc. Amer. A* **37**, 1614–1621 (2020).
20. Y. Ata and K. Kiasaleh, "Analysis of optical wireless communication links in turbulent underwater channels with wide range of water parameters," *IEEE Trans. Veh. Technol.* **72**, 6363–6374 (2023).
21. E. B. Kraus and J. A. Businger, *Atmosphere-Ocean Interaction*, Oxford University (1994).
22. M. Denny, *Air and Water: The Biology and Physics of Life's Media*, Princeton University Press, Princeton, NJ (1993).

**Shideh Tayebnaimi** (student member, IEEE) received her BSc and MSc degrees in electrical engineering in 2012 and 2015, respectively. She is currently pursuing her PhD in electrical and computer engineering at The University of Texas at Dallas. Her research focuses on improving the performance of optical communication systems, with interests in optical turbulence and wave propagation in atmospheric and underwater environments, advancing technologies in Free Space Optics (FSO) and Underwater Wireless Optical Communication (UWOC).

**Kamran Kiasaleh** received his BS (cum laude), MS, and PhD degrees all in electrical engineering from the Communications Sciences Institute at the University of Southern California in 1981, 1982, and 1986, respectively. He is a full professor and associate department head in the Department of Electrical Engineering at the University of Texas at Dallas. His research interests include optical beam propagation through turbulent/unknown media, tissue optics, and optical communication systems.